\documentclass[a4paper,11pt]{article}
\pdfoutput=1 

\usepackage{jcappub} 
\usepackage[T1]{fontenc} 

\title{Explaining Excess Dipole in NVSS Data Using Superhorizon Perturbation}

\author[a,b]{Kaustav K. Das,}
\author[a]{Kishan Sankharva,}
\author[a]{Pankaj Jain}

\affiliation[a]{Department of Physics, Indian Institute of Technology Kanpur, Uttar Pradesh - 208016}
\affiliation[b]{Division of Physics, Mathematics, and Astronomy, California Institute of Technology,\\ Pasadena,
CA 91125, USA}

\emailAdd{kdas@astro.caltech.edu}
\emailAdd{kishans@iitk.ac.in}
\emailAdd{pkjain@iitk.ac.in}

\abstract{Many observations in recent times have shown evidence against the standard assumption of isotropy in the Big Bang model. Introducing a superhorizon scalar metric perturbation has been able to explain some of these anomalies. In this work, we probe the net velocity arising due to the perturbation. We find that this extra component does not contribute to the CMB dipole amplitude while it does contribute to the dipole in large scale structures. Thus, within this model’s framework, our velocity with respect to the large scale structure is not the same as that extracted from the CMB dipole, assuming it to be of purely kinematic origin. Taking this extra velocity component into account, we study the superhorizon mode's implications for the excess dipole observed in the NRAO VLA Sky Survey (NVSS). We find that the mode can consistently explain both the CMB and NVSS observations. We also find that the model leads to small contributions to the local bulk flow and the dipole in Hubble parameter, which are consistent with observations. The model leads to several predictions which can be tested in future surveys. In particular, it implies that the observed dipole in large scale structure should be redshift dependent and should show an increase in amplitude with redshift.  We also find that the Hubble parameter should show a dipole anisotropy whose amplitude must increase with redshift in the CMB frame. Similar anisotropic behaviour is expected for the observed redshift as a function of the luminosity distance.}


\begin{document}
\maketitle
\flushbottom

\section{\textbf{Introduction}}
The most widely accepted model of cosmology is the standard $\Lambda$CDM cosmology. It explains many cosmological observations, including the cosmic microwave background (CMB) and the primordial $^4\text{He}$ abundance. The standard model's core assumption, known as the cosmological principle, is that the universe is homogeneous and isotropic on large distance scales. Within the inflationary big bang model, the universe is expected to become isotropic during the early stages of inflation \cite{wald1983}.

Recently, however, there is mounting evidence against the isotropy of the universe. Data from Wilkinson Microwave Anisotropy Probe (WMAP) has shown anomalies such as asymmetry in the CMB power spectrum \cite{dunkley2009} and alignment of the CMB quadrupole and octupole axes \citep{Costa2004}. The Planck data \cite{Planck2014} also show similar anomalies. Furthermore, anisotropy has been observed in the propagation of radio polarizations from distant galaxies \cite{Pjain1999}, and in luminosity-temperature relation of X-ray galaxy clusters \cite{X_ray_L_T_relation_anis}. Interestingly, the quadrupole-octupole alignment axis, the radio polarization dipole axis, and the CMB dipole axis roughly align along the same direction \cite{Ralston2004}, implying a possible violation of isotropy. The optical polarizations from distant quasars show alignment over large
distance scales \cite{Hutsemekers:1998}. This effect also maximizes in the direction of CMB dipole \cite{Ralston2004}.  A statistically significant dipole is also reported in the same direction in CMB fluctuations at the position of
X-ray galaxy clusters \cite{dark_flow_1, dark_flow_2, dark_flow_3, dark_flow_4, dark_flow_5, dark_flow_6, dark_flow_7}.

One way to test the isotropy is to compare our velocity relative to the CMB rest frame with that relative to the large scale structure. If the universe is isotropic, the two velocities should be identical. 
Our motion relative to the CMB frame gives rise to the CMB dipole signal. This signal has been observed to be of the order of $10^{-3}$ \cite{Kogut1993, Hinshaw2009}, and predicts our velocity to be around $369\ \rm km\ s^{-1}$ along the galactic longitude and latitude $(l,\ b) =(264^\circ,\ 48^\circ)$ \cite{Planck2018a,Planck2018b}. This local velocity is borne of the gravitational attraction of various massive structure, including the Virgo, the Great Attractor Hydra-Centaurus, Coma, Hercules, and Shapley superclusters, lying along the direction of the CMB dipole.

Our velocity relative to the large scale structure generates a dipole, known as velocity dipole or radio dipole, in the sky brightness and number count of galaxies due to aberration and Doppler effects. Various detailed maps of the local universe, including supernova catalogues, show that the nearby structure cannot account for at least 20\% of the CMB dipole value \cite{Hudson2004, Watkins2009, Watkins2014, Lavaux2010, Feldman2010, Colin2011, Feindt2013}. Thus, it is necessary to go beyond our local universe to assert if our velocity with respect to the large scale structure converges to the value predicted by CMB.

Radio astronomy provides an excellent opportunity to probe the universe at large scales. The radio sky surveys, such as the NRAO VLA Sky Survey (NVSS \cite{Condon1998}), the Sydney University Molonglo Sky Survey (SUMSS \cite{Mauch2003}), and the TIFR GMRT Sky Survey (TGSS \cite{Intema2017}) have been used to measure our velocity relative to the large scale structure. These radio surveys are relatively wide and deep, peaking at a redshift of z $\approx$ 1 \cite{Wilman2008}.  The galaxy clustering at such redshift ranges contributes very little to the dipole if we assume the $\Lambda$CDM background power spectrum. However, we expect our local motion to produce Doppler and aberration effects, which leads to a dipole in the radio galaxy distribution \cite{Ellis1984}. We do observe a dipole in these radio surveys in roughly the same direction as CMB dipole, but surprisingly the magnitude is much higher .

The NVSS is a catalogue of nearly 2 million radio galaxies covering the sky above declination of $\sim -40^\circ$. Various attempts have been made to calculate the radio dipole in the NVSS catalogue and check for convergence to the CMB dipole \cite{blake2002velocity,Singal2011, Gibelyou2012, Rubart2013, kothari2013, Tiwaripjain2015, Siewert2020}. In almost all the studies, the velocity of our local motion obtained from the NVSS dipole exceeds that obtained from the CMB dipole. In \cite{Tiwari2016}, it is shown that the NVSS dipole is at least $2.3\sigma$ greater than the CMB predicted velocity dipole. This excess dipole and the mismatch in velocity of our local motion could imply that the standard cosmological principle does not hold and that the radio galaxy distribution is anisotropic at large distance scales. 

 In analysing the data, one has to account for the incomplete sky coverage and various systematic errors. It is also important to exclude sources within the local supercluster since we are interested primarily in the cosmological signal and not local clustering. Many methods have been adopted for this purpose \cite{blake2002velocity,Tiwaripjain2015}, but the excess dipole still stays. Colin et al. (2017) \cite{Colin2017} combined the NVSS and SUMSS data to achieve a full-sky coverage. They also reported a dipole signal that is similar in direction and magnitude to that obtained in the NVSS catalogue \cite{Singal2011, Gibelyou2012, Rubart2013, kothari2013, Tiwaripjain2015}.

The TGSS catalogue \cite{Intema2017}, comprising of  0.62 million sources, also predicts a very high dipole signal \cite{Bengaly2018, Singal2019TGSS}. However, the dipole signal from TGSS disagrees not only with the CMB predictions, but also with the dipole signal from NVSS. There might be some unknown systematic error present in the TGSS data \cite{Dolfi2019,Tiwari:2019wmj}, and so, in this paper, we focus on the higher-than-expected value of the velocity dipole found in the NVSS data and explain it using a particular extension of the $\Lambda$CDM cosmology.

It has been suggested that a superhorizon perturbation may be responsible for the observed violations of statistical isotropy \cite{Gordon:2005ai,Gordon:2006,erickek2008a,erickek2008b}. Superhorizon implies that the wavelength of the perturbation is greater than the particle horizon. Ghosh (2014) \cite{Shamik2014} used the superhorizon perturbation to explain the anomaly in the NVSS data. Such a perturbation has also been invoked to explain the hemispherical power asymmetry in CMB \cite{Gordon:2005ai,Gordon:2006}. The model assumes presence of a single mode of very large wavelength which is aligned with the CMB dipole. This can be regarded as a promising toy model since it has the potential to explain several observations. A theoretical interpretation of such a mode may be provided by arguing that before inflation the Universe was not isotropic and homogeneous. This is consistent with the Big Bang paradigm in which the Universe is expected to acquire isotropy and homogeneity during the early stages of inflation \cite{1983PhRvD..28.2118W}. Based on this idea it has been proposed \cite{Rath:2013bfa} that large distance scale modes might arise during the initial stages of inflation and hence might be aligned. However a detailed application of this idea to current observations is so far lacking in the literature. In particular, it is not clear why the large distance scale modes would be aligned with the local inhomogeneities which lead to the CMB dipole.

Following the earlier work \cite{Gordon:2005ai,Gordon:2006,erickek2008a,erickek2008b,Shamik2014}, we assume an adiabatic superhorizon perturbation to explain the anomalous dipole value in the NVSS data. Unlike previous work in this area \cite{Shamik2014}, we include the following new elements: i) an extra velocity component that arises due to the superhorizon perturbation, ii) contribution to the dipole from gravitational redshift, iii) an evolving, rather than a constant, relationship between the potential and the density perturbation, and iv) the evolution of the amplitude of the perturbation in the epoch of interest which coincides with a transition from a matter-dominated universe to a dark energy-dominated universe. We investigate how well our model explains the observation data. We also make predictions on radio dipole that could be used to test the validity of the model with future surveys. Further, we check if our model is consistent with observed anisotropy in the Hubble constant and the observed bulk flow velocity.

\section{Our Velocity with respect to the Large Scale Structure} \label{sec:2}

The presence of a superhorizon scalar perturbation generates a cosmic velocity field, which leads to an additional component in our velocity with respect to the cosmic frame of rest, over and above peculiar (local) velocity. Our local velocity, which arises due to local inhomogeneities in the matter distribution, manifests itself in the CMB dipole. It is extracted by assuming that this dipole is purely kinematic. We denote this local velocity by $v_{\rm local} \hat{z}$, where we have chosen $z$-axis as the direction of the CMB dipole. The adiabatic superhorizon mode does not affect this dipole because the CMB photons get redshifted (blueshifted) as they emerge from the potential well (peak) created by the superhorizon mode, cancelling the blueshift (redshift) due to our motion induced by the mode \cite{Grishchuk1978, Turner1991, Bruni1994}. Thus, $\vec{v}_{\rm local}$ remains unaffected by the existence of this mode. In contrast, as we shall show in this paper, the superhorizon mode does affect our velocity relative to the large scale structure. In order to obtain our total velocity with respect to the large scale structure, we compute the velocity arising from the superhorizon mode and add $\vec{v}_{\rm local}$ to it. We calculate this total velocity in this section. We point out that the cancellation in CMB dipole happens only if we assume the superhorizon mode to be adiabatic. In general, if we allow isocurvature perturbations, the situation would be more complicated.

\begin{figure}
    \centering
    \includegraphics[width=\columnwidth]{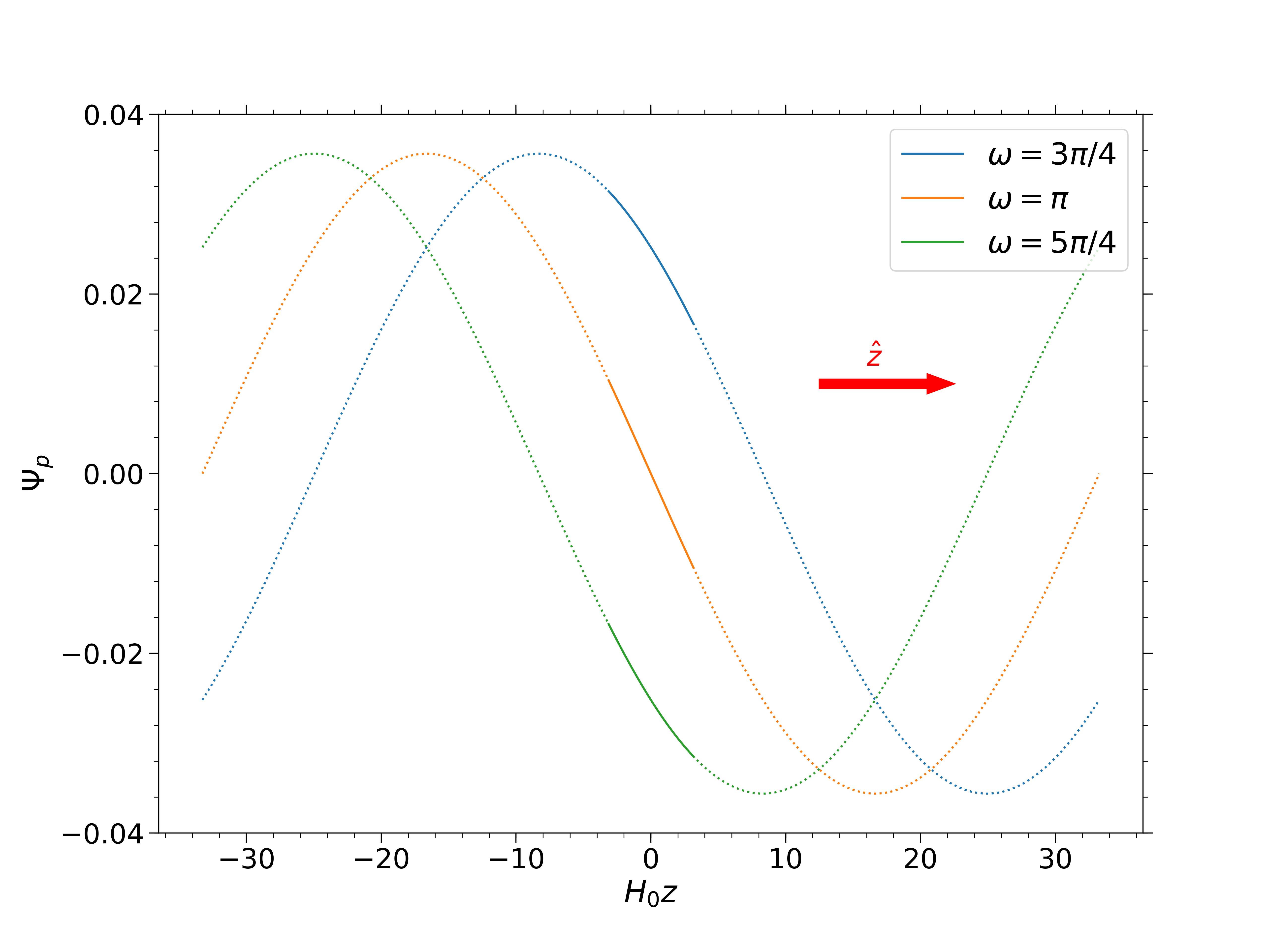}
    \caption{Primordial perturbation mode for $\alpha = 0.0356$, $\kappa = 0.0945H_0$ and $\omega = 3\pi/4,\ \pi,\ 5\pi/4$. The reason for this choice of parameter values will be explained in \S\ref{constraint}. The solid lines represent the region lying within the surface of last scattering ($z \sim 1100$). The observer is at $H_0z = 0$, where $z$ is the comoving Cartesian coordinate.}
    \label{fig:primordial_psi}
\end{figure}

We consider a scalar perturbation to the flat FLRW metric, which in the Newtonian gauge is given by ${\rm d}s^2 = -(1 + 2\Psi){\rm d}t^2 + a^2(t)(1 - 2\Phi)\delta_{ij}{\rm d}x^i{\rm d}x^j$. As the initial condition for $\Psi$, we choose a single adiabatic superhorizon mode with amplitude $\alpha$ and wavevector $\vec{k} = \kappa\hat{z}$.
\begin{gather}
    \Psi_{\rm p} = \alpha\sin(\kappa z + \omega). \label{eq:primordial_perturbation}
\end{gather}
Here, $\omega$ is a constant phase factor, and `p' in the subscript stands for primordial. The perturbation mode is shown in Fig. \ref{fig:primordial_psi} for $\omega = 3\pi/4,\ \pi,\ 5\pi/4$. The region of perturbation lying within the surface of last scattering has been shown in solid lines. Note that since the perturbation is superhorizon, the sine function does not turn over within the surface of last scattering. This form of perturbation has been motivated by a general expression used in Erickcek et al. (2008) \cite{erickek2008b}. It was also used in Ghosh (2014) \cite{Shamik2014} with $\omega=0$.
We have taken the perturbation to lie along the $z$-axis because the dipole in the NVSS data is found to lie in the direction of our velocity with respect to the CMB \cite{Singal2011}. Note that in Eq. \ref{eq:primordial_perturbation}, $z$ is the comoving Cartesian coordinate, and not redshift. Beyond this point, we use $z$ to denote redshift, and $r\cos\theta$ to represent the Cartesian coordinate $z$. Here $r$ is the comoving distance from the origin and $\theta$ is the polar angle measured with respect to the $z-$axis.

From Fig. \ref{fig:primordial_psi}, we notice that if $\omega$ is close to $\pi/2$ or $3\pi/2$, the perturbation will be symmetric about us, that is, $z = 0$. Such a perturbation mode cannot generate dipole anisotropy in the matter density contrast. However, for a wide range of other values of $\omega$ we expect a significant dipole with maximal amplitude for $\omega=\pi$. Here we have assumed that $\alpha>0$.

We use the following simplifications throughout the paper. We assume a universe without anisotropic stress. Hence, the potential $\Phi$ in the Newtonian gauge is identical to $\Psi$. Further, since the perturbation is superhorizon, we assume $\kappa/H_0 \ll 1$, where $H_0$ is the Hubble constant.
For calculations, we assume a $\Lambda$CDM model with parameter values: $\Omega_{M} = 0.3111$, $\Omega_{\Lambda} = 0.6889$, $z_{\text{eq}} = 3387$ and $H_{0}= 67.66 \text{ km s}^{-1}\ \text{Mpc}^{-1}$ (from Planck Collaboration (2018) \cite{Planck2018}). Here, $\Omega_M$ is the matter density parameter, $\Omega_{\Lambda}$ is the dark energy density parameter, and $z_{\text{eq}}$ is the redshift of matter-radiation equality. We use the natural units for speed of light ($c= 1$).

It can be shown that during the current epoch, the perturbation as a function of time is given by \cite{erickek2008b}
\begin{gather}
    \Psi(a) = \frac{9}{4}\Psi_\text{p} \Omega_M H_0^2 \frac{\widetilde{H}(a)}{a}\int_0^a \frac{\text{d}a'}{[a'\widetilde{H}(a')]^3} \label{eq:Psi_evolution}
\end{gather}
where
\begin{gather}
    \widetilde{H}^2(a) = H_0^2\bigg(\frac{\Omega_M}{a^3} + \Omega_{\Lambda}\bigg),
\end{gather}
and $a=1/(1+z)$ is the scale factor normalised to unity at the current epoch.  Eq. \ref{eq:Psi_evolution} is valid long after matter-radiation equality ($a(1+z_{eq}) \gg 1$). Since the NVSS data extends only up to $z = 2$, we use Eq. \ref{eq:Psi_evolution} for this work.

The cosmic velocity field generated by the perturbation can be computed via \cite{erickek2008b}
\begin{gather}
    \vec{v}(a,\vec{r}) = -\frac{2a^{2}}{H_{0}\Omega_{M}}\ \frac{H(a)}{H_{0}} \bigg(\frac{y}{4+3y}\bigg) \bigg [\vec{\nabla}\Psi + \frac{\text{d}}{\text{d}\ln a} \vec{\nabla}\Psi\bigg] \label{eq:cosmic_velocity_computation}
\end{gather}
where $y=a(1+z_{\text{eq}})$, and
\begin{gather}
    H^{2}(a)= H_{0}^{2} \bigg[\frac{\Omega_{M}}{a^{4}}\bigg(\frac{1}{1+z_{\text{eq}}}\bigg)+\frac{\Omega_{M}}{a^{3}}+ \Omega_{\Lambda}\bigg].
\end{gather}
Using Eq. \ref{eq:Psi_evolution} and Eq. \ref{eq:cosmic_velocity_computation}, it is straightforward to show that
\begin{gather}
    \vec{v}(z, \vec{r}) \equiv \vec{v}(z, \theta) = -\frac{9}{2}\alpha\frac{\kappa}{H_0} f(z) \cos(\kappa r(z) \cos \theta + \omega)\hat{z} \label{eq:cosmic_velocity_field}
\end{gather}
where
\begin{gather}
    f(z) = (1 + z)\frac{y}{4 + 3y}\frac{H(z)}{\widetilde{H}(z)}\bigg[\frac{H_0}{\widetilde{H}(z)} - \frac{3}{2}\Omega_MH_0^3(1+z)\int_z^{\infty} \frac{1+z'}{\widetilde{H}^3(z')}\text{d}z'\bigg], \label{eq:f}
\end{gather}
and
\begin{gather}
    r(z) = \int_0^z \frac{\text{d}z'}{H(z')}.
\end{gather}
Here, we have replaced scale factor $a$ in favour of redshift $z$.

\begin{figure}[!ht]  
	\includegraphics[width=\columnwidth]{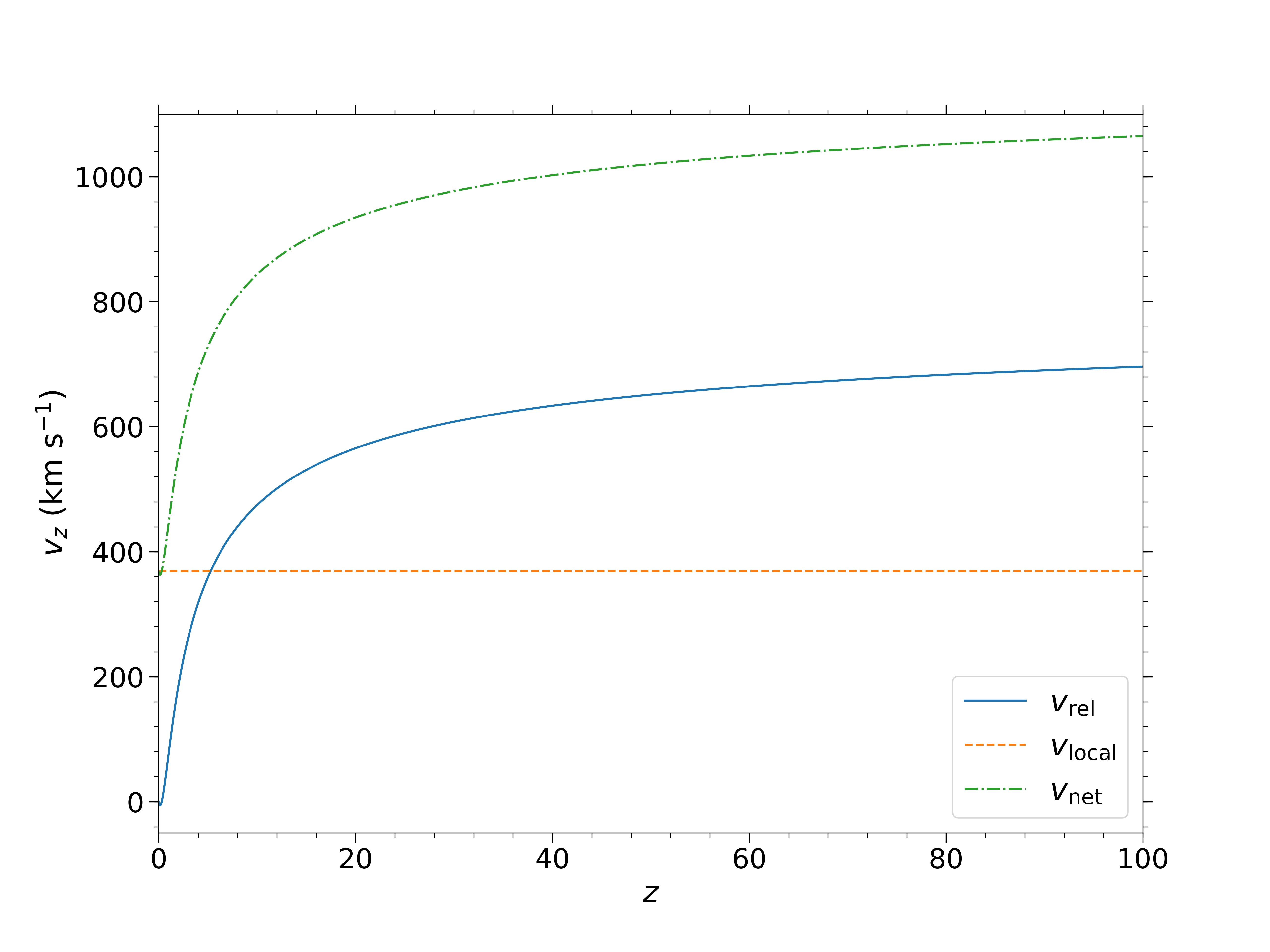}
    \caption{Lower bound on predicted velocity as a function of redshift. The blue solid line represents the velocity obtained via Eq. \ref{eq:rel_vel} using $\alpha = 0.0356$, $\kappa = 0.0945H_0$ and $\omega = \pi$. The orange dashed line represents our velocity relative to the CMB. We note that it is shown in the plot for reference and does not depend on redshift. The net velocity (Eq. \ref{eq:v_net_new}) obtained by adding the two components is shown by the dot-dash green line. The velocities are along the $z$-direction.}
    \label{fig_velplot}
\end{figure}

Our velocity relative to an object at redshift $z$ and polar angle $\theta$ can be computed by
\begin{equation} 
    \vec{v}_{\text{rel}}(z, \theta)=\vec{v}(z=0)- \vec{v}(z, \theta). 
    \end{equation}
Keeping terms only up to first order in $\kappa/H_0$, we obtain
\begin{equation}
    \vec{v}_{\rm rel}(z) = \frac{9}{2}\alpha\frac{\kappa}{H_0}[f(z) - f(0)] (\cos\omega)\ \hat{z}. \label{eq:rel_vel}
 \end{equation}
 Note that $v_{\rm rel}$ does not depend on $\theta$ at the lowest order. This is our velocity relative to an object 
 at redshift $z$ due to the superhorizon mode.
In order to obtain our net velocity with respect to the large scale structure, we add the velocity of the Sun relative to the CMB, which is given by $v_\text{local}= (369.0\ \pm\ 0.9)\  \text{km s}^{-1}$ along $(l,\ b) = (263.99^{\circ},\ 48.26^{\circ})\ \pm\ (0.14^{\circ},\ 0.03^{\circ})$ in the galactic coordinates \cite{Hinshaw2009}. Then,
\begin{gather}
    \vec{v}_{\text{net}}(z) = [{v}_{\text{rel}}(z) + v_{\text{local}}]\hat{z}. \label{eq:v_net} \\
   v_{\rm net} = \frac{9}{2}[f(z) - f(0)]\alpha\frac{\kappa}{H_0}\cos\omega + v_{\text{local}} \label{eq:v_net_new} 
\end{gather}

We plot the velocities as a function of redshift given by Eq. \ref{eq:rel_vel} and Eq. \ref{eq:v_net_new} in Figure \ref{fig_velplot} using $\alpha = 0.0356$, $\kappa = 0.0945H_0$ and $\omega = \pi$. As we shall see later in \S\ref{constraint}, these parameter values correspond to the smallest positive value of $\alpha$ that can explain the NVSS and the CMB data simultaneously. For comparison, we have also plotted the local peculiar velocity of $369$ km s$^{-1}$. Since the smallest plausible value of $\alpha$ is used, Figure \ref{fig_velplot} shows the minimum theoretical prediction for the bulk flow velocity as a function of redshift. Note that $v_{\rm rel}$ first decreases till $z = 0.15$, and then increases. This behaviour arises due to dark energy which starts to contribute at relatively small redshifts.

From Eq. \ref{eq:rel_vel}, we note that the magnitude of the velocity decreases as the magnitude of cos\ $\omega$ decreases. In particular, for $\omega = \pi/2$ and $\omega = 3\pi/2$, $\vec{v}_{\rm rel}$ will be zero at first order in $\kappa/H_0$. In general, $\omega$ may be different from $\pi$ and has to be determined by a detailed fit to data. Later, we will discuss how our results change for $\omega$ different from $\pi$.

\section{Gravitational Redshift: SW and ISW Effect due to the Superhorizon Perturbation}

The perturbation also leads to a gravitational redshift. We calculate the redshift due to the Sachs-Wolfe (SW) and the integrated Sachs-Wolfe (ISW) effect in this section. The temperature fluctuations of photons due to the SW and the ISW effects, upto first order in $\kappa$, are given by
\begin{gather}
    \bigg(\frac{\Delta T}{T}\bigg)_{\rm SW} = \Psi(z, \vec{r}) - \Psi(z=0, \vec{r}=\vec{0}) \label{eq:SW_effect}
\end{gather}
and
\begin{gather}
    \bigg(\frac{\Delta T}{T}\bigg)_{\rm ISW} = \mathcal{I}_0 \sin\omega + (\mathcal{I}_0 - \mathcal{I}_1)\vec{\kappa}\cdot\vec{r}\cos\omega, \label{eq:ISW_effect}
\end{gather}
respectively \cite{erickek2008b}. $\mathcal{I}_n$ is defined as
\begin{gather}
    \mathcal{I}_n = \frac{2\alpha}{r^n}\int_z^0 [r(z) - r(z')]^n\frac{{\rm d}g}{{\rm d}z'}{\rm d}z'.
\end{gather}
Here, we have defined $g(z)$ using $\Psi(z, \vec{r}) = g(z)\Psi_{\rm p}$, where $\Psi(z, \vec{r})$ is given by Eq. \ref{eq:Psi_evolution}.
The gravitational redshift computed from Eq. \ref{eq:SW_effect} and Eq. \ref{eq:ISW_effect}, upto first order in $\kappa$, is given by
\begin{multline}
    z_{\rm grav} = -\bigg(\frac{\Delta T}{T}\bigg)_{\rm SW+ISW} = -\alpha\sin\omega [g(0) - g(z)] + \alpha \kappa\cos\omega\cos\theta \ r(z)g(z) \\ - 2\alpha\kappa\cos\omega\cos\theta \int_0^z \frac{g(z')}{H(z')} {\rm d}z'. \label{eq:z_grav}
\end{multline}

\section{Overdensity of Matter due to the Superhorizon Perturbation\label{sec:Matter_overdensity}}

In this section, we compute the matter overdensity. In the first order perturbation theory in a flat universe with no entropy perturbations, the potential $\Psi$ and the matter overdensity $\Delta_{M}$ are related during late time via (see Equation 5.27 of Dodelson \cite{Dodelson2003})
\begin{gather}
    \Delta_M = -\frac{2a^3}{\Omega_M}\frac{H^2(a)}{H^2_0}\bigg[\Psi + \frac{\text{d}}{\text{d}\ln a}\Psi\bigg] \label{eq:Psi_Delta_relation}
\end{gather}
where we have neglected radiation and the terms of order $\kappa^2$. Using Eq. \ref{eq:Psi_evolution}, we can easily show that
\begin{gather}
    \Delta_M = -\frac{9}{2}\alpha F(z) \sin(\kappa r\cos\theta + \omega) \label{eq:DeltaM_evolution}
\end{gather}
where
\begin{gather}
    F(z) = \frac{H^2(z)}{H_0\widetilde{H}(z)}\bigg[\frac{H_0}{\widetilde{H}(z)} - \frac{3}{2}\Omega_MH_0^3(1+z)\int_z^{\infty} \frac{1+z'}{\widetilde{H}^3(z')}\text{d}z'\bigg]. \label{eq:F}
\end{gather}
The density perturbations arising from all the other modes in a homogeneous and isotropic universe can simply be added to the above result because we are working in the linear regime \cite{Gorbunov2011}. However, in the present study, we work with only the inhomogeneous term, given by Eq. \ref{eq:DeltaM_evolution}, because the contribution of the remaining modes to the dipole is expected to be negligible on the distance scale of interest.

\section{Modeling Dipole Anisotropy in Large Scale Structure}

The radio dipole observed in the large scale structure has three contributions: the kinematic dipole $D_{\rm kin}$, the gravitational dipole $D_{\rm grav}$ and the intrinsic dipole $D_{\rm int}$. The kinematic dipole arises as a result of the Doppler and aberration effects due to $\vec{v}_{\text{net}}$ -- our velocity relative to the large scale structure, while the gravitational dipole results from the Doppler effect due to the gravitational redshift $z_{\rm grav}$. On the other hand, the intrinsic dipole appears because of the inhomogeneous distribution of matter which is encoded in the overdensity $\Delta_M$. As described earlier, the superhorizon mode contributes to all -- $D_{\text{kin}}$, $D_{\rm grav}$ and $D_{\text{int}}$. Here, we determine this additional contribution besides the dipole arising due to our velocity, $v_{\text {local}}$, with respect to the CMB. Since the perturbation is assumed to be a single-mode aligned along the $z-$axis, all three contributions lie along $\hat{z}$. Hence, the observed dipole is given by
\begin{gather}
    \vec{D}_{\text{obs}} = (D_{\text{kin}} + D_{\rm grav} + D_{\text{int}})\hat{z}. 
\end{gather}

To compute the dipole, consider the flux density of the radio sources assumed to be described by a power law $S(\nu)$ (units: $\rm W\ m^{-2}\ Hz^{-1}$) depending on frequency $\nu$ as 
\begin{equation}
  S(\nu) \propto \nu^{-\alpha'}. \label{eq:S(nu)}
\end{equation}
Note that there is no relation between $\alpha$ of Eq. \ref{eq:primordial_perturbation} and $\alpha'$ of Eq. \ref{eq:S(nu)}. The integral source count per unit solid angle above a lower flux limit can also be approximated by a power law (see \cite{kothari2013})
\begin{equation}
    \frac{dN}{d\Omega} (>S) \propto S^{-x}. \label{eq:dN_dOmega_defn}
\end{equation}
Here, we take the spectral indices as $x\approx1$ and $\alpha'\approx0.75$ \cite{Gibelyou2012}. 
For an observer moving with a velocity $v = v_{\rm net}$, the Doppler shift in the frequency
is $\nu_{\rm obs} = \nu_{\rm rest}\delta$, where
\begin{equation}
    \delta \approx 1 + v_{\rm net}\cos\theta
\end{equation}
at leading order. We also need to add the contribution from the gravitational redshift. So, we have $\nu_{\rm obs} = \nu_{\rm rest}\delta_1$, where
\begin{equation}
    \delta_1 \approx 1 + v_{\rm net} \cos\theta - z_{\rm grav} .
\end{equation}
The relation between the observed and the actual flux at a fixed frequency due to the Doppler effect is then given by \cite{Ellis1984}
\begin{equation}
S_{\rm obs} = S_{\rm rest}\delta_1^{1+\alpha'}. \label{eq:S_obs_S_rest}
\end{equation}
Furthermore, the aberration effect changes the solid angle in the direction of motion according to ${\rm d}\Omega_{\rm obs} = {\rm d}\Omega_{\rm rest}\delta^{-2}$. 

Let ${\rm d}^2N_{\rm rest}$ and ${\rm d}^2N_{\rm obs}$ represent the number of sources in the bin ${\rm d}\Omega_{\rm rest}{\rm d}S_{\rm rest}$ and ${\rm d}\Omega_{\rm obs}{\rm d}S_{\rm obs}$, respectively. From Eq. \ref{eq:dN_dOmega_defn}, we find
\begin{gather}
    \frac{{\rm d}^2N_{\rm rest}}{{\rm d}\Omega_{\rm rest}{\rm d}S_{\rm rest}} = kx(S_{\rm rest})^{-1-x}
\end{gather}
where $k$ is a proportionality constant. We have ${\rm d}^2N_{\rm rest} = {\rm d}^2N_{\rm obs}$, and therefore, we obtain
\begin{equation}
    {\rm d}^2N_{\rm obs} = {\rm d}^2N_{\rm rest} = kx (S_{\rm rest})^{-1-x}\delta^2{\rm d}\Omega_{\rm obs}{\rm d}S_{\rm rest}.
\end{equation}
Substituting $S_{\rm rest}$ from Eq. \ref{eq:S_obs_S_rest}, we obtain
\begin{equation}
    {\rm d}^2N_{\rm obs} = kx(S_{\rm obs})^{-1-x} \delta^{x(1+\alpha')}_1\delta^2 {\rm d}\Omega_{\rm obs}  {\rm d}S_{\rm obs}.
\end{equation}
Integrating over $S_{\rm obs}$ from $S_{\rm low}$ to $\infty$, we get
\begin{equation}\label{eq:source_counts}
    \bigg(\frac{{\rm d}N}{{\rm d}\Omega}\bigg)_{\rm obs} = k(S_{\rm low})^{-x}\delta^2\delta^{x(1+\alpha')}_1 = \bigg(\frac{{\rm d}N}{{\rm d}\Omega}\bigg)_{\rm rest} \delta^2\delta^{x(1+\alpha')}_1.
\end{equation}

We point out that each of the terms $({\rm d}N/{\rm d}\Omega)_{\rm obs}$, $({\rm d}N/{\rm d}\Omega)_{\rm rest}$ and $\delta^2\delta^{x(1+\alpha')}_1$ can be a function of redshift $z$ in addition to $\theta$ and $\phi$ (the polar angles).
We can express $(\text{d}N/\text{d}\Omega)_{\text{rest}}$ as
\begin{gather}
    \bigg(\frac{\text{d}N}{\text{d}\Omega}\bigg)_{\text{rest}} = \bigg(\frac{\text{d}N}{\text{d}\Omega}\bigg)_{\text{rest}}^{(0)} + \Delta \bigg(\frac{\text{d}N}{\text{d}\Omega}\bigg)_{\text{rest}}
\end{gather}
where $(\text{d}N/\text{d}\Omega)_{\text{rest}}^{(0)}$ is the isotropic zeroth order term and $\Delta(\text{d}N/\text{d}\Omega)_{\text{rest}}$ is a perturbation to the zeroth order term. Note that $(\text{d}N/\text{d}\Omega)_{\text{rest}}^{(0)} \propto \rho_M$ and $\Delta(\text{d}N/\text{d}\Omega)_{\text{rest}} \propto \delta\rho_M(\theta, \phi)$. Hence, we can write
\begin{gather}
    \frac{(\text{d}N/\text{d}\Omega)_{\text{rest}}}{(\text{d}N/\text{d}\Omega)_{\text{rest}}^{(0)}} = 1 + \Delta_M b(z).
\end{gather}
$\Delta_M = \delta\rho_M/\rho_M$ is the matter overdensity, and $b(z)$ is the galaxy bias factor. To compute the overdensity of galaxies from the matter overdensity, we need to multiply $\Delta_M$ by the galaxy bias factor $b(z)$ (see \cite{Nusser2015}). Dividing Eq. \ref{eq:source_counts} by $(\text{d}N/\text{d}\Omega)_{\text{rest}}^{(0)}$, we get
\begin{gather}
    \frac{(\text{d}N/\text{d}\Omega)_{\text{obs}}}{(\text{d}N/\text{d}\Omega)_{\text{rest}}^{(0)}} = (1 + \Delta_M b(z))\delta^2\delta^{x(1+\alpha')}_1.
\end{gather}
We expand $\delta^2\delta^{x(1+\alpha')}_1$ up to first order in $\cos\theta$
\begin{gather}
    \delta^2\delta^{x(1+\alpha')}_1 = 1 + v_{\rm net}[2+x(1+\alpha')]\cos\theta - z_{\rm grav}x(1+\alpha')
\end{gather}
Therefore,
\begin{gather}
    \frac{(\text{d}N/\text{d}\Omega)_{\text{obs}}}{(\text{d}N/\text{d}\Omega)_{\text{int}}^{(0)}} = 1 + v_{\rm net}[2+x(1+\alpha')]\cos\theta - z_{\rm grav}x(1+\alpha')  + \Delta_M b(z) \label{eq:source_counts_final_eq}
\end{gather}
where we have retained terms upto first order in $\kappa/H_0$ only. The coefficient of $\cos\theta$ in the second, third and fourth terms on the right hand side of the above equation leads to $D_{\text{kin}}$, $D_{\rm grav}$ and $D_{\text{int}}$, respectively. 

To compute the dipole amplitude, we calculate the spherical harmonic coefficients $a_{lm}$ which, for a general function $f(\theta, \phi)$, are given by
\begin{gather}
    a_{lm} = \int f(\theta, \phi)Y^{*}_{lm}(\theta, \phi) \text{d}\Omega.
\end{gather}
The $l=1$ components represent the dipole. Due to our choice of coordinate system, $a_{11} = 0 = a_{1,-1}$. Thus, we only need to evaluate $a_{10}$. The dipole amplitude $D$ is computed using \cite{Shamik2014, Gibelyou2012}
\begin{gather}
    D = a_{10}\sqrt{\frac{3}{4\pi}}. \label{eq:dipole_amplitude}
\end{gather}
The dipole, thus obtained, has a dependence on redshift. In order to get the projected dipole value $\Bar{D}$, we multiply $D$ with the normalised galaxy distribution function $w(z)$ of the survey under study and integrate over the redshift range ($z_1$ to $z_2$) over which the survey data extends.
\begin{gather}
    \Bar{D} = \int_{z_1}^{z_2} D(z)w(z) \text{d}z. \label{eq:projected_dipole}
\end{gather}

\subsection{Kinematic Dipole}

The kinematic dipole results from the second term in Eq. \ref{eq:source_counts_final_eq}. Using Eq. \ref{eq:v_net_new} and Eq. \ref{eq:dipole_amplitude}, we obtain
\begin{gather} \label{eq:4.12}
    D_{\rm kin}(z) = \frac{9}{2}[2 + x(1+\alpha')][f(z) - f(0)]\alpha\frac{\kappa}{H_0}\cos\omega + v_{\rm local}[2+x(1+\alpha')].
\end{gather}
This leads to the following value of projected dipole
\begin{equation}
    \Bar{D}_{\text{kin}} = \mathcal{A}_1(z_1,\ z_2) \alpha\frac{\kappa}{H_{0}}\cos\omega + \mathcal{B}
\end{equation}
where
\begin{gather}
    \mathcal{A}_1(z_1,\ z_2) = \frac{9}{2}[2+x(1+\alpha')] \bigg[\int_{z_1}^{z_2} f(z) w(z) \text{d}z - f(0)\bigg]
\end{gather}
and
\begin{gather}
    \mathcal{B} = v_{\text{local}}[2+x(1+\alpha')].
\end{gather}

\subsection{Gravitational Dipole}

The third term on RHS of Eq. \ref{eq:source_counts_final_eq} gives rise to the gravitational dipole. Using Eq. \ref{eq:z_grav} and Eq. \ref{eq:dipole_amplitude}, we find
\begin{gather} \label{eq:grav_dipole}
    D_{\rm grav}(z) =  -x(1+\alpha')\bigg[\alpha \kappa\cos\omega\ r(z)g(z) - 2\alpha\kappa\cos\omega \int_0^z \frac{g(z')}{H(z')} {\rm d}z'\bigg].
\end{gather}
Note that the first term in Eq. \ref{eq:z_grav} is a monopole term and does not appear in the above equation. The projected dipole is given by
\begin{gather}
    \Bar{D}_{\text{grav}} = \mathcal{A}_2(z_1,\ z_2)\alpha\frac{\kappa}{H_{0}}\cos\omega
\end{gather}
where
\begin{gather}
    \mathcal{A}_2(z_1,\ z_2) = -x(1+\alpha')H_0\int_{z_1}^{z_2}\bigg[r(z)g(z) - 2 \int_0^z \frac{g(z')}{H(z')} {\rm d}z'\bigg]w(z){\rm d}z.
\end{gather}

\subsection{Intrinsic Dipole}

The intrinsic dipole arises due to the matter overdensity $\Delta_M$. Expanding Eq. \ref{eq:DeltaM_evolution} up to first order in $\kappa$, and using Eq. \ref{eq:dipole_amplitude}, we obtain
\begin{gather} \label{eq:4.16}
    D_{\rm int}(z) = -\frac{9}{2}r(z)b(z)F(z)\alpha\kappa\cos\omega.
\end{gather}
The corresponding projected dipole is
\begin{equation}
    \Bar{D}_{\text{int}} = \mathcal{C}(z_1,\ z_2)\alpha\frac{\kappa}{H_0}\cos\omega, \label{eq:intrinsic_matter}
\end{equation}
where
\begin{gather}\label{C_def}
    \mathcal{C}(z_1,\ z_2) = -\frac{9}{2}H_0 \int_{z_1}^{z_2} r(z)F(z)w(z)b(z) \text{d}z.
\end{gather}
Adding the three contributions, the net projected dipole is given by
\begin{gather}
    \vec{D}_{\text{obs}} = \bigg[\big\{\mathcal{A}_1(z_1,\ z_2) + \mathcal{A}_2(z_1,\ z_2) + \mathcal{C}(z_1,\ z_2)\big\}\alpha\frac{\kappa}{H_0} \cos\omega + \mathcal{B}\bigg]\hat{z}. \label{eq:dipole_amplitude_final}
\end{gather}
This is the predicted value of dipole in the presence of the superhorizon mode.








\begin{table}[!tbp]
\centering
\begin{tabular}{|c|c|}
\hline
& \\
\textrm{Author} &
\textrm{\hspace{8mm}Dipole Amplitude $D_{\rm obs}$\hspace{8mm}} \\
& \\
\hline 
& \\
Tiwari et al. (Number count)\cite{kothari2013} & $0.0151 \pm\ 0.0030$  \\
\hspace{8mm}Tiwari et al. (Sky Brightness) \cite{kothari2013} \hspace{8mm} & $0.0166 \pm\ 0.0031$   \\
Singal \cite{Singal2011} & $0.019 \pm\ 0.004$ \\
Gibelyou and Huterer \cite{Gibelyou2012} & $0.027 \pm\ 0.005$  \\
Rubart and Schwarz \cite{Rubart2013} & $0.018 \pm\ 0.006$ \\
Bengaly et al. \cite{Bengaly2018} &  $0.0234 \pm\ 0.0039$ \\
& \\

\hline
\end{tabular}
\caption{\label{tab:dipole_values}Net observed dipole amplitudes.}
\end{table}

\section{Comparing the Predicted Dipole with the Observed NVSS Dipole} \label{constraint}
In this section, we compare the dipole value predicted by our model with the observed values. The observed NVSS dipole amplitudes are given in Table \ref{tab:dipole_values}. The estimated dipole directions are not the same for the different methods used in these papers \cite{Singal2011,kothari2013,Gibelyou2012,Rubart2013,Bengaly_Jr__2016} and also differ to some extent from the CMB dipole. Here we shall ignore these small variations in direction and focus only on the amplitude. The spread in dipole values arise due to different methodology employed for extracting the dipole. In particular, Tiwari et al. \cite{kothari2013} obtain lower value since they eliminate nearby sources which may bias the cosmological signal.

To calculate the predicted dipole, we use the galaxy bias factor $b(z) = 0.33 z^2 + 0.85z + 1.6$ \cite{Nusser2015} and the following normalized galaxy distribution function of the NVSS catalogue \cite{Tiwari2016}
\begin{gather}
    w(z)=\mathcal{N} z^{0.74} \exp\bigg[-\bigg(\frac{z}{0.71}\bigg)^{1.06}\bigg]. \label{eq:galaxy_dist}
\end{gather}
Here, $\mathcal{N}$ is a normalization constant.

\begin{figure}[!ht]
	\includegraphics[width=1.0\columnwidth]{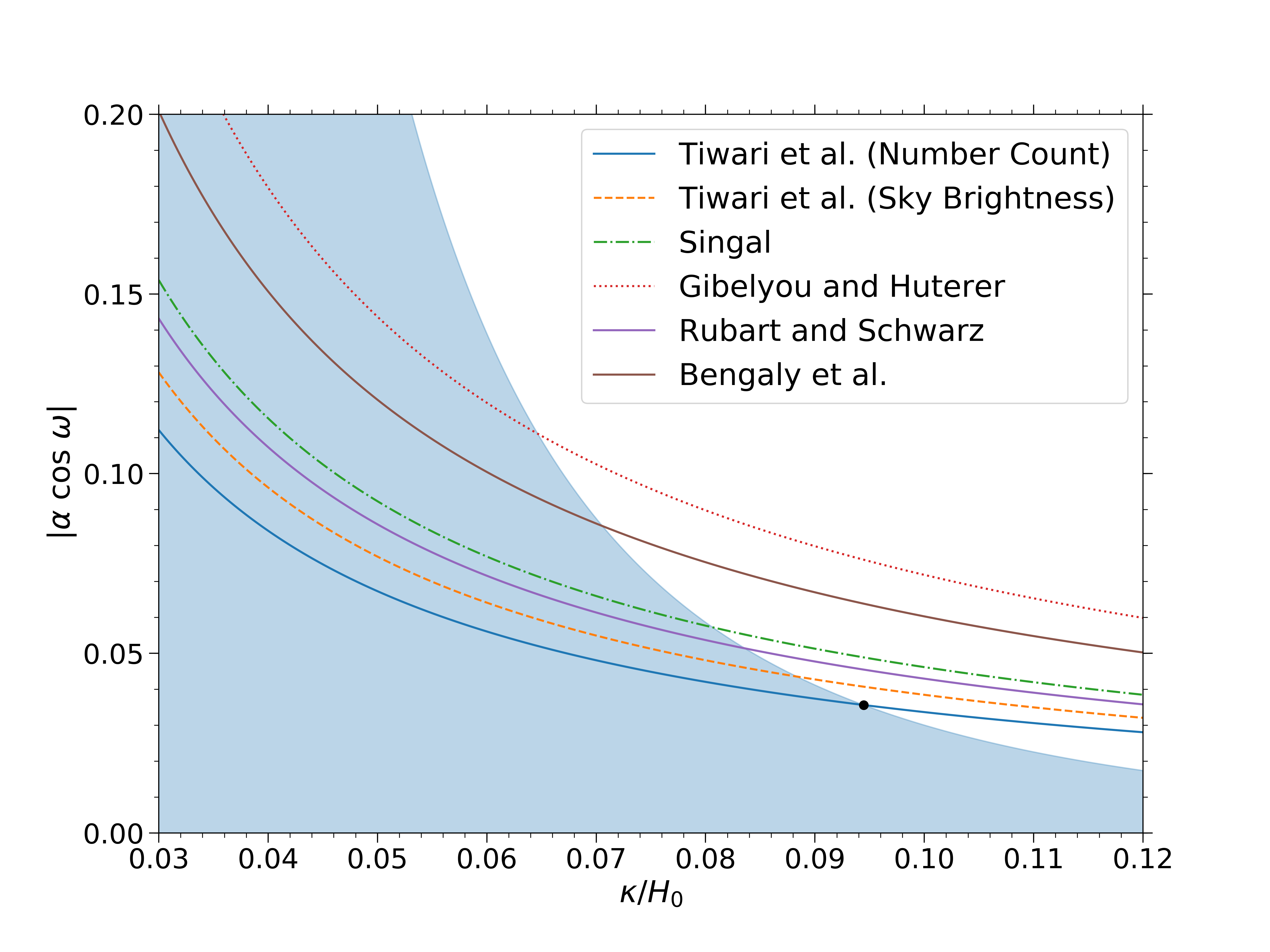}
    \caption{Plot of $|\alpha\cos\omega|$ versus $\kappa/H_0$ for the observed NVSS dipole amplitudes. The blue-shaded part denotes the region of parameter space that satisfies the CMB constraint (Eq. \ref{eqnn:34}). The black dot at the point $(|\alpha\cos\omega|,\ \kappa/H_0) = (0.0356,\ 0.0945)$ represents the minimum value of $\alpha$ required to satisfy both the NVSS and the CMB observations.} 
    \label{fig1}
\end{figure}

\begin{figure*}[!ht] 
	\includegraphics[width=1.0\textwidth,height=1\textheight,keepaspectratio=true]{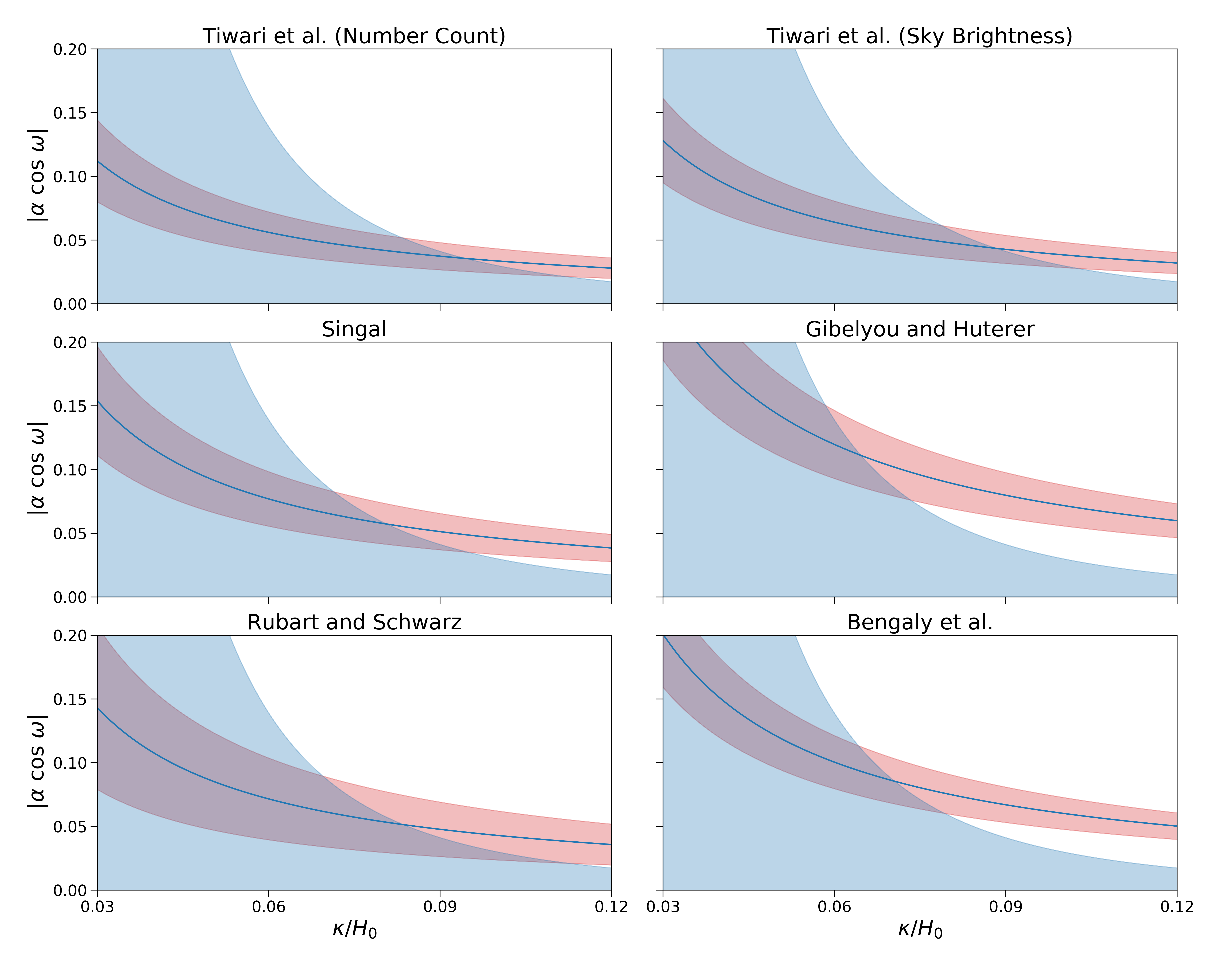}
    \caption{The $1\sigma$ confidence contours in $|\alpha\cos\omega|$ versus $\kappa/H_0$ space plotted using the dipole amplitudes and errors given in Table \ref{tab:dipole_values}. The blue-shaded part denotes the region of parameter space that satisfies the CMB constraint (Eq. \ref{eqnn:34}). }
    \label{fig_errors}
\end{figure*}

The NVSS data extends over the redshift range $z_1=0$ to $z_2=2$. Hence, we obtain
\begin{equation}
    \alpha\cos\omega = \frac{H_0}{\kappa}\frac{D_{\text{obs}} - \mathcal{B}}{\mathcal{A}_1(0,\ 2) + \mathcal{A}_2(0,\ 2) + \mathcal{C}(0,\ 2)}
\end{equation}
from Eq. \ref{eq:dipole_amplitude_final}. In Figure \ref{fig1}, we plot $|\alpha\cos\omega|$ as a function of $\kappa/H_0$ for the six values of $D_{\rm obs}$ given in Table \ref{tab:dipole_values}. 
The blue-shaded part denotes the region of parameter space that satisfies the CMB constraint
\begin{equation}
\label{eqnn:34}
    |\kappa^3 \alpha \cos\omega| \leq 2.99 \times 10^{-5} H_{0}^{3}.
\end{equation}
This constraint is obtained by demanding that the CMB multipole moment $a_{30}$ produced by the model satisfies  $|a_{30}|<3\sqrt{C_3}$, where $C_3$ is the power for $l=3$ \cite{erickek2008b}.
The errors for the predicted values of $\alpha$ are shown in Figure \ref{fig_errors}. 

We find that there exist model parameter values that fall within the region allowed by the CMB constraint while also explaining the NVSS dipole. Thus, we find that the model provides an explanation for the excess NVSS dipole. The black dot in Figure \ref{fig1} at the point $(|\alpha\cos\omega|,\ \kappa/H_0) = (0.0356,\ 0.0945)$ represents the minimum value of $\alpha$ that satisfies both the CMB and the NVSS constraints. Henceforth, we shall use the parameter values corresponding to this black point to make predictions and check for the consistency of the model. We notice that we can explain the data for a wide range of values of $\omega$ with $\alpha$ taking the smallest value for $\omega=\pi$.

\section{Testing Consistency with Hubble Dipole and Bulk Flow Observations}
 \label{hub_Cal}

\begin{figure}[!ht] 
	\includegraphics[width=\columnwidth]{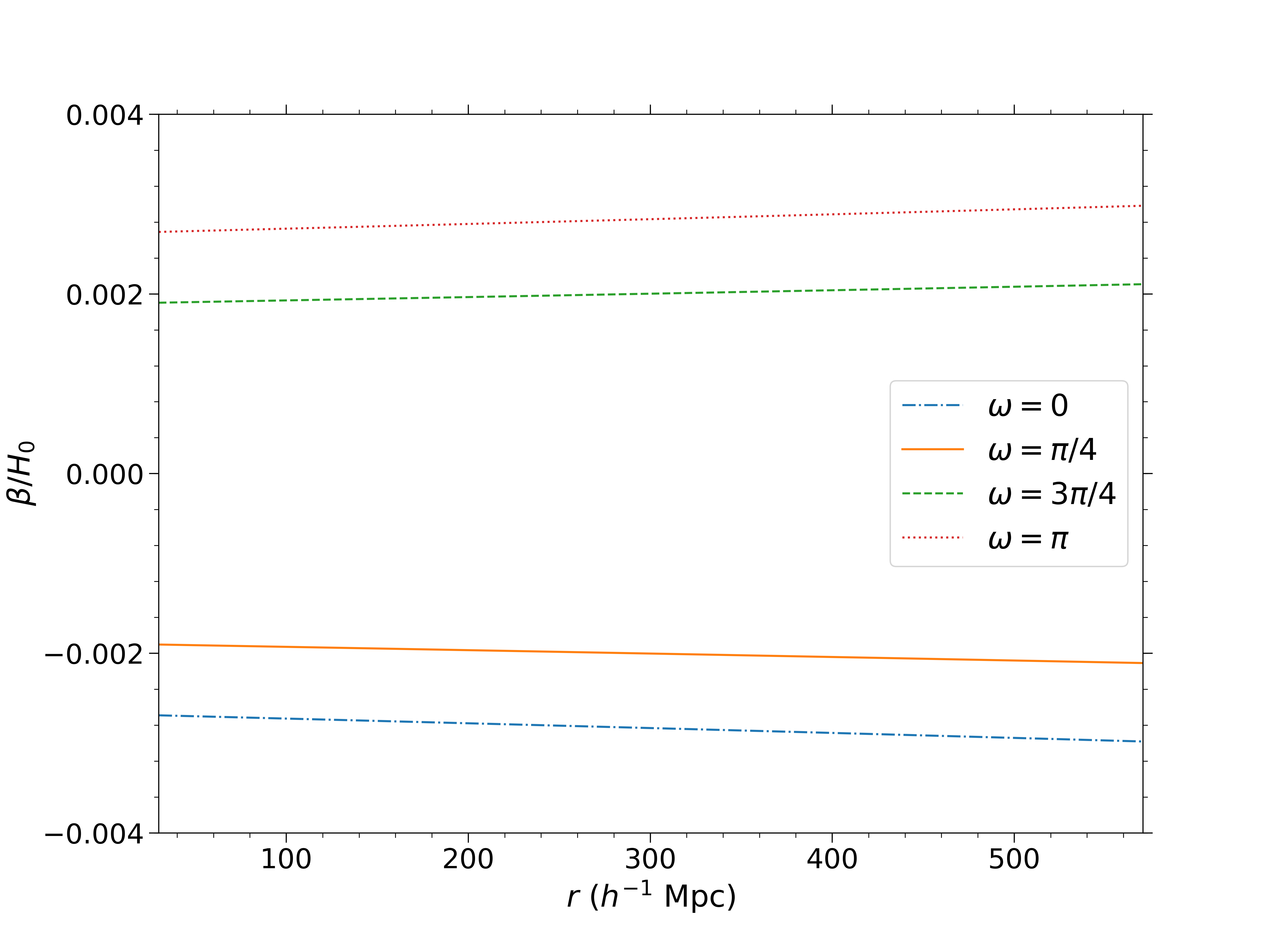}
    \caption{The predicted lower bound of Hubble constant dipole $\beta(z)/H_0$ in the CMB rest frame. $r$ is the comoving distance and $H_0$ is the direction-averaged Hubble constant during the current epoch. The parameter values used are:  $(\alpha,\ \kappa/H_0) = (0.0356,\ 0.0945)$ for $\omega=0$, $\pi/4$, $3\pi/4$, $\pi$.}
    \label{fig_betabyh0}
\end{figure}

In this section, we do a basic check for the consistency of our model with observed data for parameters that are possible manifestations of the superhorizon perturbation. The superhorizon perturbation affects the redshift of distant sources due to the Doppler and gravitational effects. This leads to a dipole anisotropy in the Hubble parameter. We determine the magnitude of Hubble dipole using our model in order to confirm that it is within observational limits.

Let $z$ be the cosmological redshift in the absence of the superhorizon perturbation. The perturbation gives rise to $z_{\rm Doppler}$, the Doppler redshift due to our velocity relative to the large scale structure, and $z_{\rm grav}$, the gravitational redshift due to potential wells and peaks. The observed redshift $z_{\rm obs}$ can then be written as
\begin{equation}
    1+z_{\rm obs} = (1+z)(1+z_{\rm Doppler})(1+z_{\rm grav}).
\end{equation}
The additional contributions, $z_{\rm Doppler}$ and $z_{\rm grav}$, introduce an anisotropy in the observed redshift.  We model this as $z_{\rm obs} = \Bar{z} + \gamma \cos\theta$. Using $z_{\rm Doppler} = - v_{\rm net} \cos\theta$, and $z_{\rm grav}$ from Eq. \ref{eq:z_grav}, we obtain
\begin{equation} \label{eq:gamma}
    \Bar{z} = z + \alpha\sin\omega (1 + z)[g(z) - g(0)]
\end{equation}    
    and
    \begin{equation}
    \gamma = \alpha\kappa\cos\omega(1+z)\bigg[r(z)g(z) - 2\int_0^z \frac{g(z')}{H(z')} {\rm d}z'\bigg] - v_{\rm net}(1+z)[1+\alpha\sin\omega\{g(z)-g(0)\}]  \label{eq:z_obs_anisotropy}
\end{equation}
up to first order in $\kappa$.

Anisotropy in the observed redshift translates to an anisotropy in the Hubble constant inferred from the observed redshift. We model this anisotropy as $H^{\rm obs}_0 = \Bar{H}^{\rm obs}_0 + \beta\cos\theta$. In the local universe, we have $z_{\rm obs} = H^{\rm obs}_0 r$. Comparing the coefficients of $\cos\theta$, we get
\begin{equation}
    \beta(z) =  \frac{\gamma}{r(z)}.
\end{equation}
$\beta(z)$ is the dipole component of the Hubble constant inferred from observed redshift. It lies along the $z$-direction. We plot the predicted $\beta(z)/H_0$ in the CMB frame as a function of the comoving distance in Figure \ref{fig_betabyh0}. For the CMB frame, we use $v_{\rm local} = 0$ in Eq. \ref{eq:v_net_new}. The parameter values used are: $(\alpha,\ \kappa/H_0) = (0.0356,\ 0.0945)$ for $\omega=0$, $\pi/4$, $3\pi/4$, $\pi$. We note that to explain the excess NVSS dipole, we need negative value for $\cos\omega$, so that $v_{\rm rel}$ is in the direction ($+\hat{z}$) of the local motion (see Eq \ref{eq:rel_vel}). This gives us a positive kinematic dipole which generates a blueshift. However, for this to happen, we must have an over-density in that direction which creates a potential well, redshifting the photons from $+\hat{z}$. In the case of LSS, the redshift due to potential well dominates and this leads to a positive Hubble anisotropy for us. This explains why we have positive values of $\beta$ for $\omega = 3\pi/4,\ \pi$ in Figure \ref{fig_betabyh0}. One can similarly explain the trends for other values of $\cos\omega$. 

\subsection{Hubble Constant Dipole in COMPOSITE Sample}

The COMPOSITE sample of Watkins, Feldman and Hudson \cite{Watkins2009, Feldman2010} is one of the largest available data set of peculiar velocities of galaxies that extend upto $\sim 100h^{-1}$ Mpc. Wiltshire et al. (2013) \cite{Wiltshire2013} use this sample to characterize the anisotropy in the Hubble flow, adopting an analysis methodology independent of any cosmological model assumptions except that a suitably defined average linear Hubble law exists.

\begin{figure}[!ht] 
	\includegraphics[width=\columnwidth]{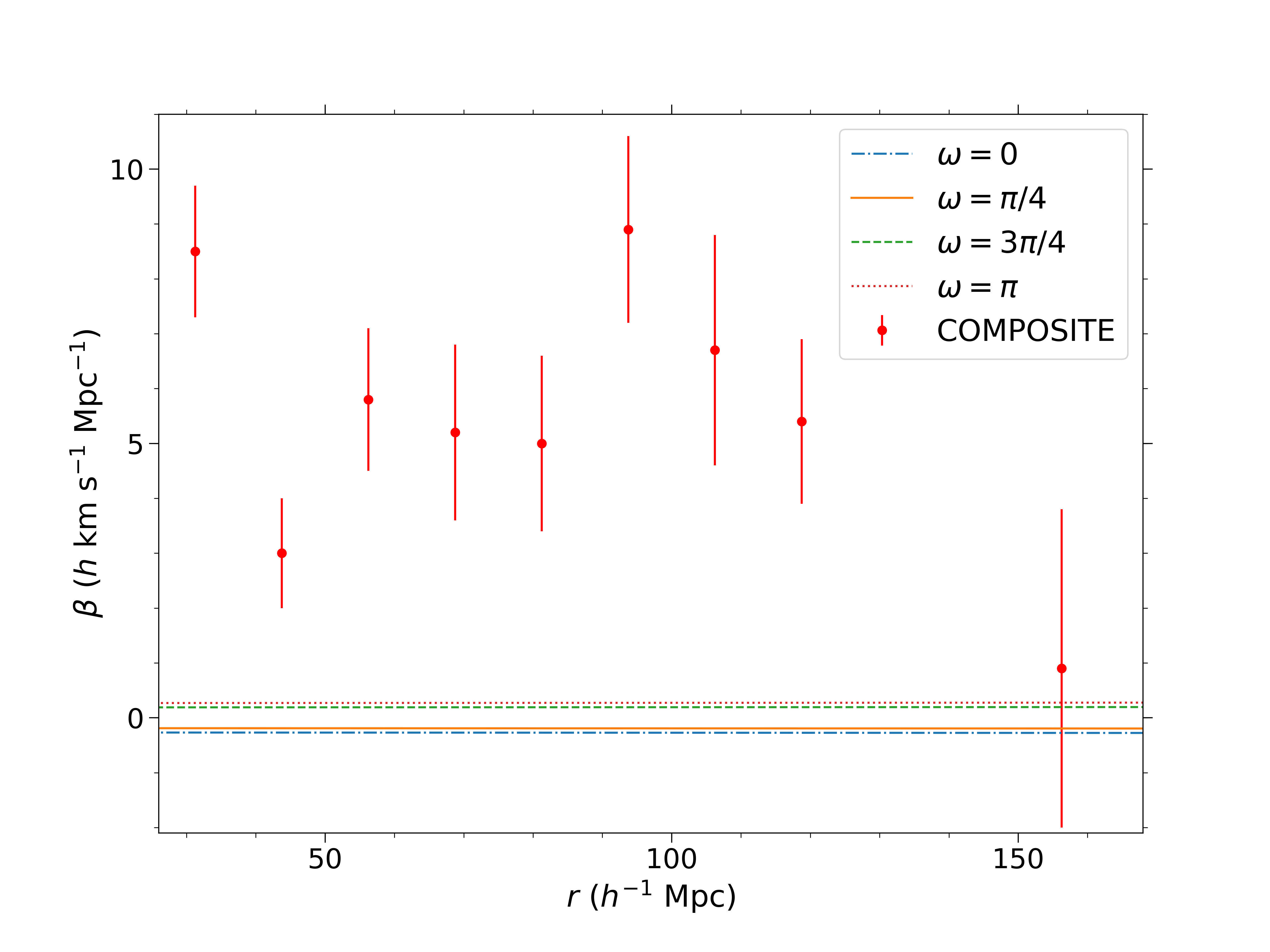}
    \caption{The predicted lower bound on the Hubble constant dipole calculated using the minimum value of amplitude of the perturbation mode that obeys both the NVSS and the CMB constraints obtained in \S\ref{constraint}:  $(\alpha,\ \kappa/H_0) = (0.0356,\ 0.0945)$ for $\omega=0$, $\pi/4$, $3\pi/4$, $\pi$. The dipole values and errors obtained by Wiltshire et al. (2013) \cite{Wiltshire2013}, using the COMPOSITE Survey, in the CMB rest frame are shown by the red dots and the red vertical error bars. Here $r$ is the comoving distance.}
    \label{fig_composite}
\end{figure}

They fit a dipole Hubble law in independent radial shells. In Figure \ref{fig_composite}, we show the lower bound on the predicted Hubble constant dipole, along with the observed value obtained by Wiltshire et al. (2013) \cite{Wiltshire2013}, in the CMB rest frame. We have used $\alpha = 0.0356$ and $\kappa = 0.0945H_0$ (the smallest required value of $\alpha$ obtained in \S\ref{constraint}). We note that the direction of the dipole obtained in Wiltshire et al. (2013) \cite{Wiltshire2013} has a high variance.

We find that our lower bound prediction is consistent with the observation since the observed dipole in all the radial shells is greater than the minimum value predicted by the model. The excess observed dipole can be attributed to local effects. We postpone a more detailed analysis of the Hubble constant anisotropy using different data sets, rest frames, and analysis methods, including samples at higher redshifts, to future research.

\subsection{Bulk Flow Observations}

The gravitational pull of large scale structure gives rise to peculiar velocities of galaxies. It is preferable to study the bulk velocity, that is, the mean peculiar velocity of a large volume of space, to understand the distribution of matter. Averaging peculiar velocity over a large volume removes various non-linear effects affecting the small scales. There have been many attempts to deduce the bulk velocity in the local universe using different samples with redshift ranges varying from $z<0.05$ to $z<0.2$. The obtained bulk flow velocities vary from $292\pm 96$ km s$^{-1}$ to $188 \pm 120$ km s$^{-1}$. We refer to \cite{Bengaly_Jr__2016, Colin2011, Dai2011, Rathaus2013, Feindt2013, Mathews2016, Appleby2015} for more details on the samples and methodology used to obtain these values.

In our model, the bulk flow velocity is given by $v_{\rm net}$, defined in Eq. \ref{eq:v_net}. We obtain a $v_{\rm net}$ of about 350 km s$^{-1}$ using $\alpha$ = 0.12, $\kappa$ = 0.065$H_0$ and $\omega = \pi$ (this set of parameter values also satisfy the CMB constraint). Hence we obtain a value which is a little smaller than local velocity of 369 $\rm km\ s^{-1}$ expected from CMB dipole. This deviation arises due to the negative value of $v_{\rm rel}$ for low redshifts, as is evident from Figure \ref{fig_velplot}. 
The observed bulk flow velocity is found to be much lower than the local velocity of 369 $\rm km\ s^{-1}$ and may be attributed to local effects. Hence we do not violate the observational limits on this parameter.

\section{Signatures of the Model to be Tested in Future}

Our model makes several predictions which can be tested in future. Here, we list some predictions for radio and optical surveys. These primarily use the fact that the predicted dipole anisotropy is redshift dependent.

\subsection{Dipole Distribution in Galaxy Surveys}

\begin{figure}[!ht] 
	\includegraphics[width=\columnwidth]{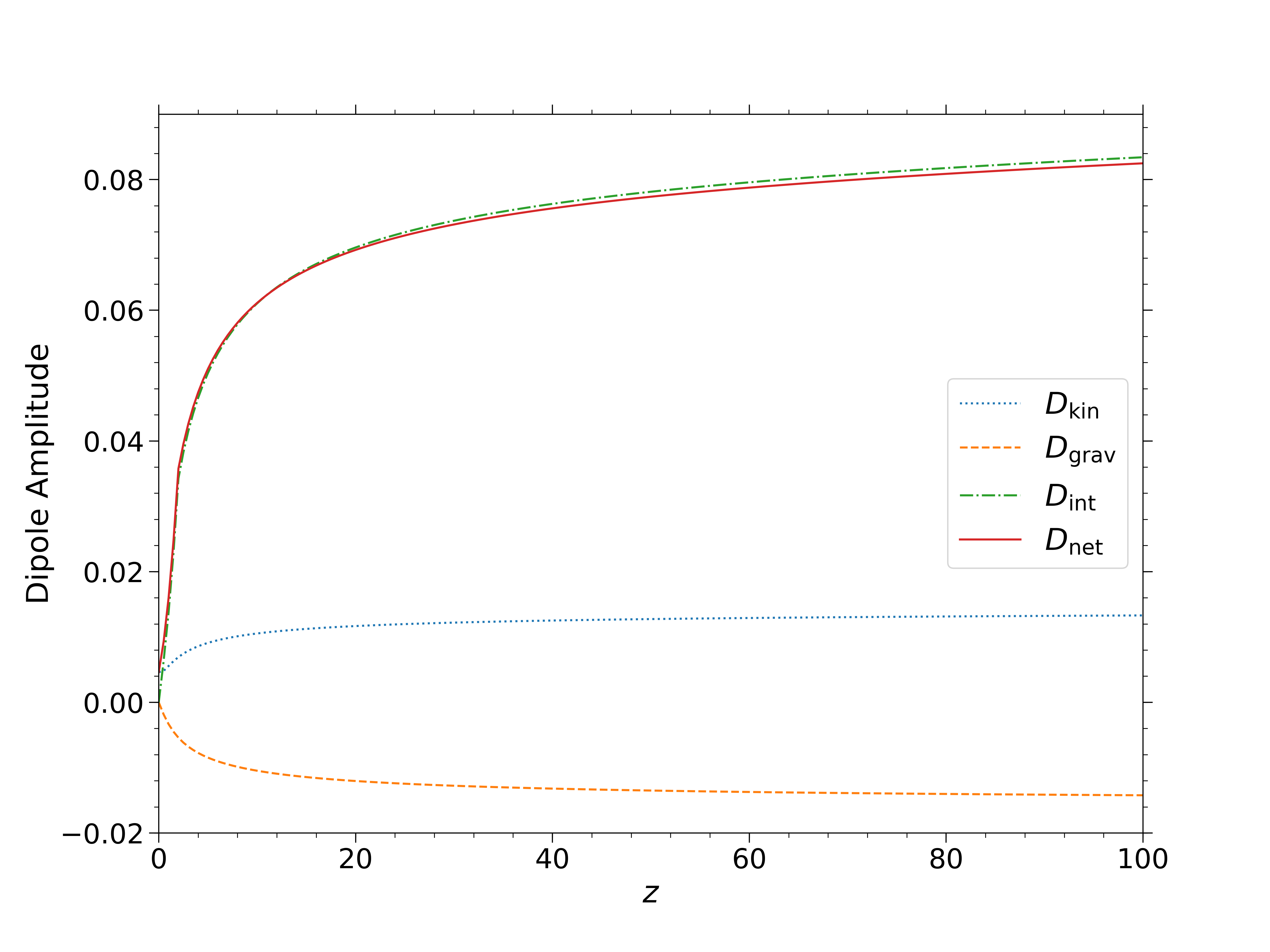}
    \caption{The predicted dipole distribution with redshift assuming the minimum amplitude of the perturbation mode that obeys both the NVSS and the CMB constraints obtained in \S\ref{constraint}: $(\alpha,\ \kappa/H_0,\ \omega) = (0.0356,\ 0.0945,\ \pi)$.}
    \label{fig_dip_distribution}
\end{figure}

An interesting property of our model is that the dipole has a significant redshift dependence. Therefore, the observed dipole value also depends on the redshift range one is looking at. The net dipole distribution as a function of redshift is $D_{\rm kin}$ + $D_{\rm grav}$ + $D_{\rm int}$, with $D_{\rm kin}$, $D_{\rm grav}$ and $D_{\rm int}$ as defined in Eq. \ref{eq:4.12}, Eq. \ref{eq:grav_dipole} and Eq. \ref{eq:4.16}, respectively. Figure \ref{fig_dip_distribution} shows the plot of the dipole distribution with redshift for $\alpha = 0.0356$, $\kappa = 0.0945 H_0$ and $\omega = \pi$ (the smallest required value of $\alpha$ obtained in \S\ref{constraint}).

We also obtain the minimum expected projected dipole amplitudes in three redshift bins, viz., $z = 0.0-0.5$, $0.5-1.0$ and $1.0-2.0$. We use Eq. \ref{eq:dipole_amplitude_final} with $z_1$ and $z_2$ as the minimum and maximum redshift values of a bin, respectively. Here too, we use the smallest value of the perturbation amplitude that obeys both the NVSS and the CMB constraints. These predicted minimum dipole amplitudes have been tabulated in Table \ref{Tablebins}. This prediction can be tested in future radio and optical surveys.

\begin{table}[tbp]  
\centering
\begin{tabular}{|c|c|}
\hline
& \\
\textrm{\hspace{8mm}Redshift bin\hspace{8mm}} &
\textrm{\hspace{8mm}Dipole Amplitude $D_{\rm 0}$\hspace{8mm}} \\
& \\
\hline
& \\
0.0 - 0.5  & 0.0072   \\
0.5 - 1.0  & 0.0124 \\
1.0 - 2.0 & 0.0235\\
& \\
\hline

\end{tabular}
\caption{\label{Tablebins} Projected dipole prediction for various redshift bins assuming the minimum amplitude of the perturbation mode that obeys both the NVSS and the CMB constraints obtained in \S\ref{constraint}: $(\alpha,\ \kappa/H_0,\ \omega) = (0.0356,\ 0.0945,\ \pi)$.}
\end{table}

\subsection{Radio Dipole Prediction for Higher Redshifts}

Another test for our model would be to verify the prediction for the minimum expected radio dipole for a survey depth of z $\sim$ 3 through future surveys. We use Eq. \ref{eq:dipole_amplitude_final}, assuming that Eq. \ref{eq:galaxy_dist} is valid till $z=3$, to obtain the predicted value. Using the smallest possible value for the amplitude of the perturbation, we find that the minimum expected dipole would be 0.0184.
Note that for predicting the dipole at even higher redshifts, the suppression factor $g(a) = \Psi(a)/\Psi_p$, given by Eq. \ref{eq:Psi_evolution}, has to be replaced by the numerical solution to Equation 11 in Erickcek et al. (2008) \cite{erickek2008b} which we do not pursue in the current paper. However we may continue to use this equation in order to get a qualitative idea about the behaviour at larger redshifts. We also assume that the galaxy bias saturates beyond a certain redshift of z $\sim$ 3. We find, using $\omega=\pi$, that the dipole amplitude increases rapidly up to $z \sim 8$ to a value of about 0.06 and then continues to increase very slowly beyond this. This increase even at large redshifts is expected since the mode is superhorizon with $\kappa = 0.0945H_0$. Thus, even for very large redshifts of $z \sim 1000$, the value of $\kappa r$ is smaller than $\pi/2$ and the sinusoidal perturbation does not turn over (see Figure \ref{fig:primordial_psi}).

\subsection{Predictions for Anisotropy in the Observed Redshift}

\begin{figure}[!ht] 
	\includegraphics[width=\columnwidth]{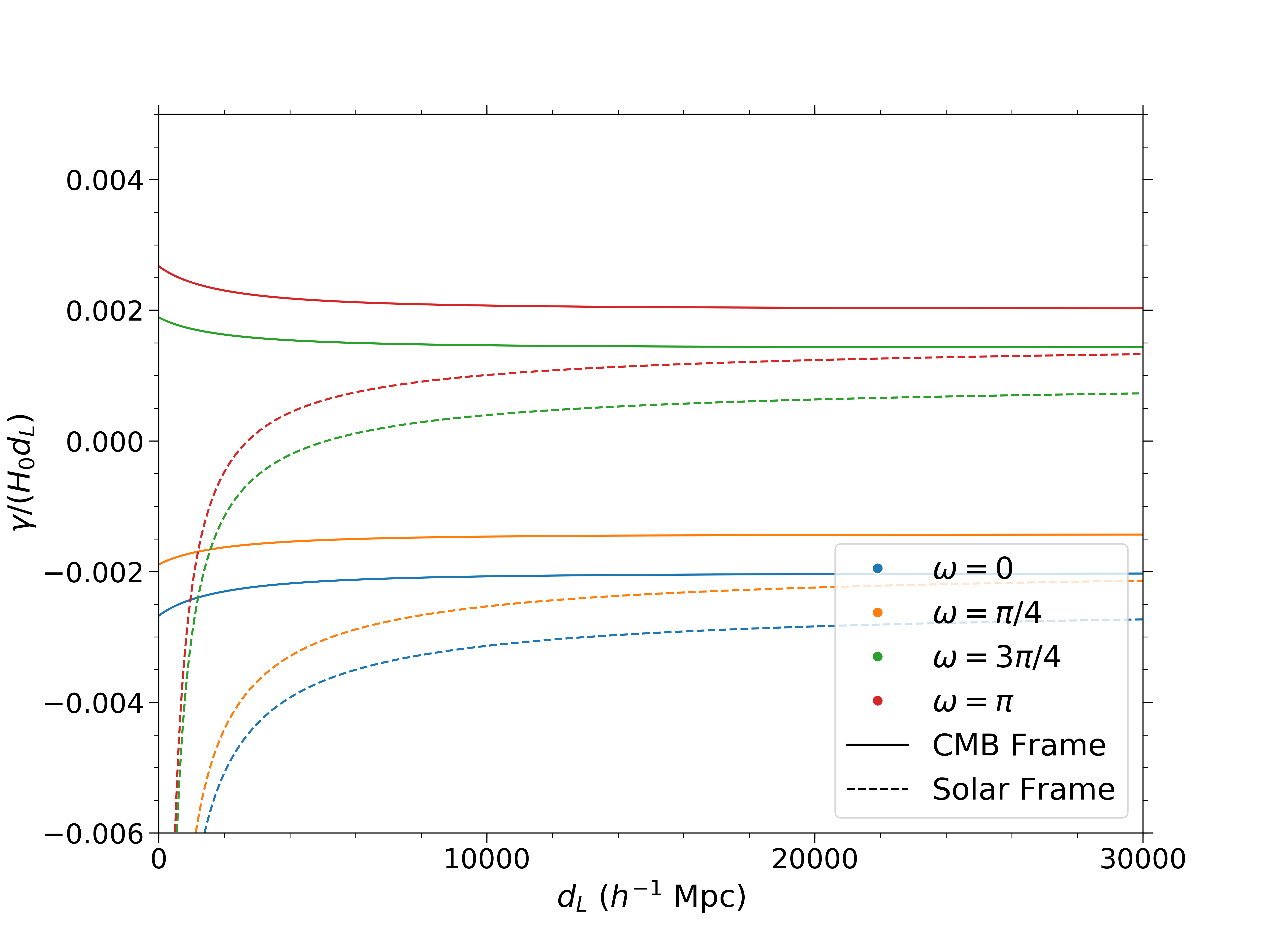}
    \caption{The predicted dipole anisotropy $\gamma/(H_0d_L)$ in the observed redshift as a function of the luminosity distance $d_L$. We use the minimum amplitude of the perturbation mode that obeys both the NVSS and the CMB constraints obtained in \S\ref{constraint}: $(\alpha,\ \kappa/H_0) = (0.0356,\ 0.0945)$ for $\omega = 0,\ \pi/4,\ 3\pi/4,\ \pi$. We note from Eq. \ref{eq:z_obs_anisotropy} that for $z\rightarrow0$, $\gamma\rightarrow0$ in the CMB frame ($v_{\rm local} = 0$), while $\gamma\rightarrow\ \sim-0.001$ in the solar frame ($v_{\rm local} \neq 0$).}
    \label{fig:z_anis}
\end{figure}

As seen in $\S$ \ref{hub_Cal}, the superhorizon perturbation introduces an anisotropy in the observed redshift, which can be modelled as $z_{\rm obs} = \Bar{z} + \gamma\cos\theta$, with $\gamma$ given by Eq. \ref{eq:z_obs_anisotropy}. We can test for this anisotropy in the present and future supernova and large scale structure surveys. In Figure \ref{fig:z_anis}, we plot $\gamma/(H_0d_L)$ as a function of the luminosity distance $d_L$ in the solar as well as the CMB frame for four different values of $\omega$, viz., $\omega = 0$, $\pi/4$, $3\pi/4$ and $\pi$. The values of $\alpha$ and $\kappa$ used are 0.0356 and 0.0945$H_0$, respectively. We observe that the dipole anisotropy as a function of the luminosity distance saturates at large distances in both the CMB and the solar frame.

\section{Conclusion}

In this paper, we have performed a detailed study of the implications of a superhorizon mode proposed to explain some cosmological observations that appear to show deviation from isotropy. In particular, we show that this mode predicts an extra velocity component of the solar system with respect to the large scale structure. This extra velocity component cancels out in the case of CMB dipole but not for large scale structure dipole. This leads to the interesting implication that, within this model's framework, our velocity with respect to the large scale structure is not the same as that extracted from the CMB dipole. The observed value of the dipole anisotropy with respect to large scale structure, however, depends not only on the relative velocity but also on the anisotropic potential induced by the mode and the gravitational redshift induced by this potential. Considering all these contributions, we obtain the parameter range that satisfies both the NVSS observation and the CMB constraints. Recently, the presence of an anomalously high dipole in the sky distribution of 1.36 million quasars using the Wide-field Infrared Survey Explorer (WISE) was reported by Secrest et al. (2020) \cite{Secrest2020} . This greatly increases the confidence of the excess dipole as WISE is systematically independent from ground-based radio observations. We note that using our model, we can also account for the dipole value of 0.01554 reported in their work which is similar to that reported using NVSS.

We also check if our model is consistent with observations of the dipole anisotropy in Hubble constant and the bulk flow velocity in the local universe. The observed Hubble constant may display a dipole anisotropy due to local contributions. Our model predicts a small anisotropy in this parameter which is much smaller than observations. The observed bulk flow velocities  show considerable spread but are generally found to be smaller than our velocity with respect to the CMB. We find that our model predicts a velocity of about 350 km s$^{-1}$, i.e. a velocity slightly smaller than that obtained from the CMB dipole. In contrast, the observations suggest a much larger deviations. Hence we are consistent both with Hubble dipole and bulk flow observations.

We show that the superhorizon mode makes several interesting predictions which can be tested in future surveys. In particular, it leads to a remarkable prediction that the bulk flow velocity and the dipole in large scale structure should increase with redshift. This increase is rapid for small redshifts, and it progressively slows down at large redshifts. The behaviour is expected since the mode is superhorizon. Similar anisotropic behaviour, which increases with redshift, is predicted for several other observables. In particular, we have demonstrated this for the Hubble parameter dipole and for the observed redshift dipole as a function of distance in the CMB frame. It will be exciting to test these predictions in future observations.


\acknowledgments

We thank Shamik Ghosh for useful comments. We also thank Prabhakar Tiwari for a very useful input. We would like to thank the anonymous referee for useful comments. We acknowledge funding from the Science and Engineering Research Board (SERB), Government of India, grant number EMR/2016/004070.


\bibliographystyle{JHEP}
\bibliography{ref1} 



\end{document}